\documentclass[entropy,article,accept,moreauthors,pdftex,10pt,a4paper]{mdpi}

\usepackage{hyperref}
\usepackage{xcolor}
\usepackage{soul}

\firstpage{1}
\articlenumber{142}
\doinum{10.3390/e20020142}
\pubvolume{20}
\pubyear{2018}
\copyrightyear{2018}
\history{Received: 19 January 2018; Accepted: 16 February 2018; Published: 22 February 2018}
\graphicspath{{./figures/}}

\Title{A Simple and Adaptive Dispersion Regression Model for Count Data}

\Author{Hadeel~S.~Klakattawi~\textsuperscript{1}\href{https://orcid.org/0000-0001-6617-2081}{\orcidicon}, Veronica~Vinciotti~\textsuperscript{2}\href{https://orcid.org/0000-0002-2625-7977}{\orcidicon} and Keming~Yu~\textsuperscript{2}\textsuperscript{,}*}
\AuthorNames{Hadeel S. Klakattawi, Veronica Vinciotti and Keming Yu}

\address{%
$^{1}$ \quad Department of Statistics, Faculty of Science, King Abdulaziz University, Jeddah  {21589}, Saudi Arabia; hklakattawi@kau.edu.sa\\%%Please add zipcode
$^{2}$ \quad Department of Mathematics, Brunel University London, Uxbridge UB8 3PH, UK; Veronica.Vinciotti@brunel.ac.uk}

\corres{Correspondence: keming.yu@brunel.ac.uk, {Tel.: +44-1895-266128}}%Please add Tel. for correspondence author

\abstract{Regression for count data is widely performed by models such as Poisson, negative binomial (NB) and zero-inflated regression. A challenge often faced by practitioners is the selection of the right model to take into account dispersion, which typically occurs in count datasets. It is highly desirable to have a unified model that can automatically adapt to the underlying dispersion and that can be easily implemented in practice. In this paper, a discrete Weibull regression model is shown to be able to adapt in a simple way to different types of dispersions relative to Poisson regression: overdispersion, underdispersion and covariate-specific dispersion. Maximum likelihood can be used for efficient parameter estimation. The description of the model, parameter inference and model diagnostics is accompanied by simulated and real data analyses.}


\begin{document}
%%%%%%%%%%%%%%%%%%%%%%%%%%%%%%%%%%%%%%%%%%
\section{Introduction}

Count data, which refers to the number of times an item or an event occurs within a fixed period of time,
commonly arises in many fields. Indeed, examples of count data include the number of heart attacks or the number of hospitalisation days in medical studies, the number of students absent during a period of time in education studies, or  the number of times parents perpetrate domestic violence against their child in social science investigations. There is now a great deal of interest in the literature on  investigating the relationship between a count response variable and other variables: for example,  how the education level of parents can affect the incidence of domestic violence against their children. Methods to address these questions fall in the general area of regression analysis of count data (see~\cite{cameron2013regression,hilbe2014modeling} among others).

Classical regression models for count data belong to the family of generalised linear models \citep{nelder1972generalized} such as  Poisson regression, which models the conditional mean of the counts as a linear regression on a set of covariates through the log link function.
Although Poisson regression is fundamental to the regression analysis of count data, it is often of limited use for real data, because of its property of an~equal mean and variance.
Real data usually feature overdispersion relative to Poisson regression, or~the opposite case of underdispersion. Thus, accounting for overdispersion and underdispersion when modelling count data is essential,
and failing to cope with these features of the data can lead to biased parameter estimates and thus false conclusions and decisions.

%%Please confirm if a new paragraph starts from here.
{Negative Binomial} (NB) regression is widely considered as the  default choice for data that are overdispersed relative to Poisson regression. However, NB regression may not be the best choice for
power-law data with long tails, or for highly skewed data with an excessive number of zeros, because of the rare occurrence of non-zero events. These often require the application of zero-inflated and hurdle models. 
In addition, NB regression cannot deal with data that are underdispersed relative to Poisson regression. There have been some attempts to extend Poisson regression-based models to underdispersion, such as generalised Poisson (GP) regression \citep{efron1986double,famoye1993restricted}, {COM}% Please define.
--Poisson regression \citep{sellers2008flexible} or hyper-Poisson regression models \citep{saez2013hyper}. However, these models are all modifications of a Poisson model and have been shown to be rather complex and computationally expensive in practice.
In~this paper, we introduce the discrete Weibull (DW) distribution to the regression modelling of count data and present the DW regression model.
The motivation behind considering the DW distribution \citep{nakagawa1975discrete} stems from the vital role played by the continuous Weibull distribution in {survival analysis} and failure time studies. However, few contributions can be found in the literature on statistical inference and applications using this distribution,  aside from some early work on parameter estimation \cite{khan1989estimating,kulasekera1994approximate} and some limited use  in applied contexts: for example, Refs.~\cite{englehardt2011discrete,englehardt2012methods}, who showed that the counts of living microbes (pathogens) in water are highly skewed and can be efficiently modelled using a DW distribution. In contrast to this, this paper shows a number of desirable features of this distribution, which are particularly appealing
within a regression context.
Specifically, this simple count regression model can capture different levels of dispersion adaptively, which is a challenge faced by existing count regression models. Moreover, we show how the DW model can capture power-law behaviour and high skewness in the underlying distributions.

Section~\ref{sec:dw} provides a review and description of the DW distribution and its properties. Section~\ref{sec:disp} illustrates the ability of a DW distribution to model data that are both overdispersed and underdispersed relative to Poisson regression. The DW regression model is introduced in Section \ref{sec:dwreg} to investigate the relationship between a count response and a set of covariates. Section \ref{sec:natuhan}  {shows the ability of the DW regression model to handle cases of mixed levels of dispersion.} This model is applied to a number of real datasets in Section \ref{sec:realdata}.

%%%%%%%%%%%%%%%%%%%%%%%%%%%%%%%%%%%%%%%%%%

\section{Discrete Weibull Distribution} \label{sec:dw}

%%%%%%%%%%%%%%%%%%%%%%%%%%%%%%%%%%%%%%%%%%

\subsection{The Distribution}
\begin{figure}[H]
	\centering
	\includegraphics[scale=0.9]{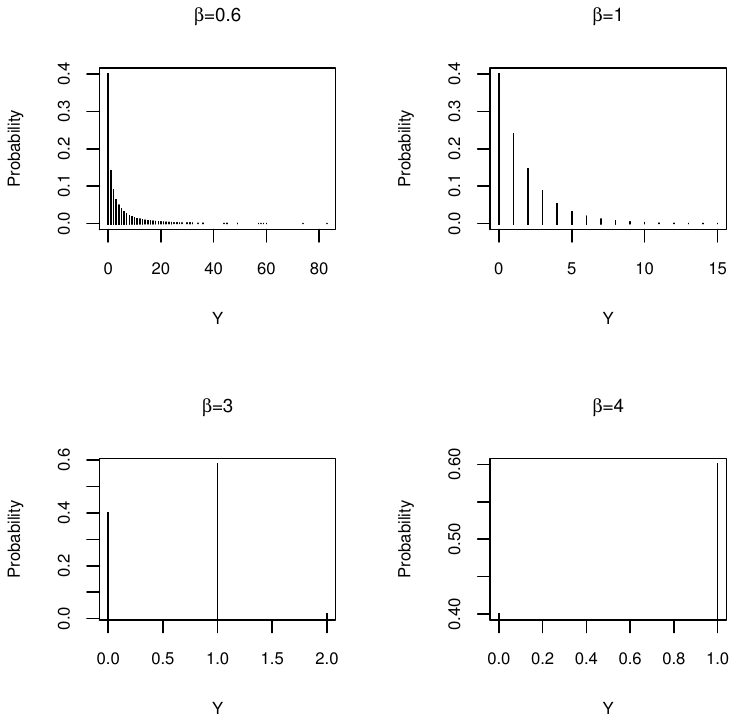}
	\caption{The effect of $ \beta $ on the  discrete Weibull {(DW) probability mass function with $ q=0.6 $.}}
	\label{fig:specialcaseofaDWdistribution}
\end{figure}

If $ Y $ follows a (type 1) DW distribution \citep{nakagawa1975discrete}, then the cumulative distribution function of $ Y $ is given~by
\begin{equation}
\label{eq:cdf discrete weibull}
F(y;q,\beta)
=
\begin{cases}
1-q^{(y+1)^{\beta}} & \mbox{\text{for} $y=0,1,2,3, \ldots$} \\
0 & \mbox{otherwise}
\end{cases}
\end{equation}
and its probability mass function is given by
\begin{equation}
\label{eq:pmf discrete weibull}
f(y;q,\beta)
=
\begin{cases}
q^{y^{\beta}}
-
q^{(y+1)^{\beta}} & \mbox{\text{for} $y=0,1,2,3, \ldots$} \\
0 & \mbox{otherwise}
\end{cases}
\end{equation}
with the parameters $ 0<q<1 $ and $ \beta>0 $.  {Because $ f(0)=1-q $, the parameter $ q $ is the probability of obtaining a non-zero response.} We refer later to this distribution as $ DW(q,\beta) $.
The distribution is connected to other well-known distributions. In particular, the following:
\begin{itemize}[leftmargin=*,labelsep=5.8mm]
	\item The discrete Rayleigh distribution in \cite{roy2004discrete} is a special case of a DW distribution with $ \beta=2 $ and $q=\theta$.
	\item The geometric distribution is a special case of a DW distribution, with $ \beta=1 $ and $q=1-p$. Moreover, for the geometric distribution, the variance is always greater than its mean. Therefore, a DW distribution with $ \beta=1$ is a case of overdispersion relative to Poisson regression, regardless of the value of $q$. In particular, when $ \beta=1 $ and $ q=e^{-\lambda} $, the distribution is the discrete exponential distribution introduced by \cite{sato1999new}.
	\item $ \beta $ can be considered as controlling the range of values of the variable. In other words, this parameter controls the skewness of the DW distribution.  {In order to show this}, Figure~\ref{fig:specialcaseofaDWdistribution}   {plots the probability mass functions} for a fixed parameter $q$ and different values of $\beta$. The plot shows how the  {the probability of $0$ stays constant, while the tail of the distribution becomes increasingly longer as $ \beta  \longrightarrow 0$, and the distribution approaches a Bernoulli distribution with probability $q$ as~$ \beta \rightarrow \infty $.}
\end{itemize}

%%%%%%%%%%%%%%%%%%%%%%%%%%%%%%%%%%%%%%%%%%

\subsection{Moments  and Quantiles}
The first two moments of a DW distribution are given by
\begin{eqnarray}
\label{eq:expectedDW}
E(Y)&=\mu&=\sum_{y=1}^{\infty} q^{y^{\beta}}\\
E(Y^{2})&=&2\sum_{y=1}^{\infty} y q^{y^{\beta}} - E(Y) \nonumber
\end{eqnarray}

From these, the variance for a DW distribution is given by
\begin{eqnarray}
\label{eq:varianceDW}
Var(Y)=\sigma^2=2\sum_{y=1}^{\infty} y q^{y^{\beta}} -\mu-\mu^2
\end{eqnarray}
for which there are no closed-form expressions, although numerical approximations can be obtained on a truncated support \citep{barbiero2013package}.  Equations~(\ref{eq:expectedDW}) and (\ref{eq:varianceDW}) show that both $ E(Y)>Var(Y) $ and  $ E(Y)<Var(Y) $  {are generally possible, making the DW distribution suitable both for overdispersion and underdispersion.}

A nice property of the DW distribution is that  its $\tau$th ($0<\tau<1$) quantile, that is, the smallest value of $ y $ for which $F(y) \ge \tau$, has a closed-form expression, given by
\begin{equation}
Q(\tau)= \Big\lceil \left(\dfrac{\log(1-\tau)}{\log(q)}\right)^{\frac{1}{\beta}}-1\Big\rceil
\label{eq:quantileDW}
\end{equation}
for $ \tau \geq 1-q$. This is in contrast to Poisson and NB regression, which do not have a closed-form expression for the quantiles.

%%%%%%%%%%%%%%%%%%%%%%%%%%%%%%%%%%%%%%%%%%

\subsection{Parameter Estimation}
Given a sample $ y_{1}, y_{2}, \ldots, y_{n} $ from a DW distribution, the log-likelihood can be written as
\begin{equation*}
\ell=
\sum\limits_{i=1}^{n}
\log
\left(
q^{y_{i}^{\beta}}
-
q^{(y_{i}+1)^{\beta}}
\right)
\end{equation*}
from which the maximum likelihood estimators (MLEs) of $ q $ and $ \beta $ can be easily obtained by directly maximising this log-likelihood using any standard nonlinear optimisation tool.

%%%%%%%%%%%%%%%%%%%%%%%%%%%%%%%%%%%%%%%%%%

\section{DW Accounts for Different Types of Dispersion} \label{sec:disp}
In this section, we discuss a property of the DW distribution that is particularly advantageous as a model for count data. Dispersion in count data is formally defined in relation to a specified model being fitted to the data \citep{cameron2013regression,hilbe2014modeling}. In particular, we let
\begin{equation}
\label{eq:varratio}
\text{VR}=\dfrac{\text{observed variance}}{\text{theoretical variance}}
\end{equation}

Thus the {variance ratio (VR)} is the ratio between the observed  variance from the data and the theoretical variance from the model. Then the data are said to be overdispersed/equidispersed/underdispersed relative to the fitted model if the observed variance is larger/equal/smaller than the theoretical variance specified by the model, respectively. 
% We remove all the italic for VR (including in the equation 6) to be consistent throughout text, please confirm.
It is common to refer to dispersion relative to Poisson regression. In this case, the variance of the model is estimated by the sample mean. Thus, overdispersion/equidispersion/underdispersion relative to Poisson regression refers to cases in which the sample variance is larger/equal/smaller than the sample mean, respectively. Because the theoretical variance of a NB regression is always greater than its mean, as $\sigma^2=\mu+\dfrac{1}{k} \mu^{2}$ for $k>0$, the NB regression model is the natural choice for data that are overdispersed relative to Poisson regression. However, crucially, NB regression cannot handle underdispersed data.

\begin{figure}[H]
	\centering
	\includegraphics[width=.48\linewidth]{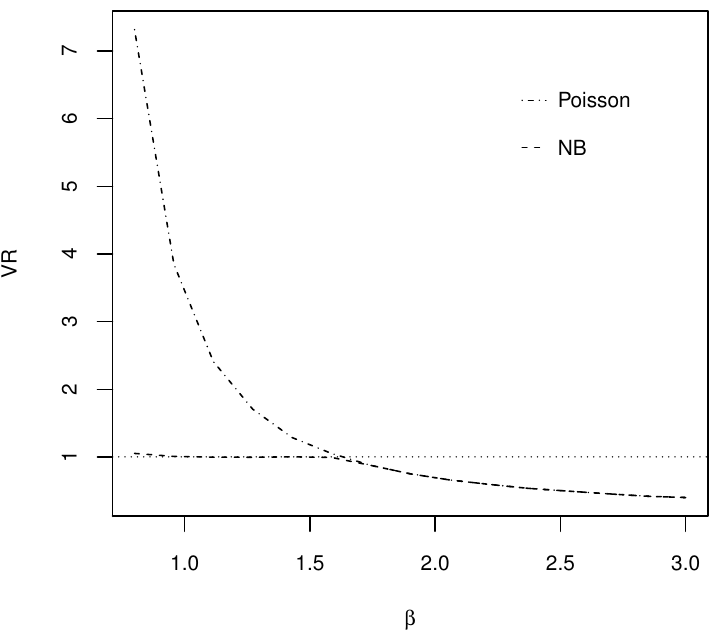}
	\includegraphics[width=.48\linewidth]{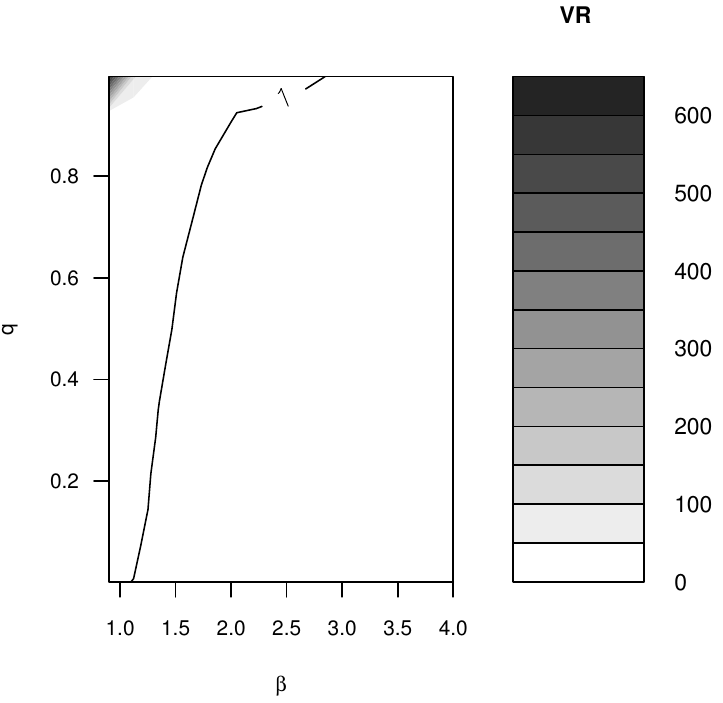}
	\caption{Ratio of observed and theoretical variance of data simulated by $DW(q,\beta)$. Left: $q=0.7$; data fitted by Poisson and negative binomial (NB) regression. Right: a range of $q$ and $\beta$ values; data fitted by Poisson regression.  {The area below 1 corresponds to cases of underdispersion relative to Poisson regression, whereas the area above 1 corresponds to cases of overdispersion.}}
	\label{fig:observed_and_fitted_variances}
\end{figure}

In contrast to this, we show how a DW distribution can handle data that are both overdispersed and underdispersed relative to Poisson regression. In particular,  Figure~\ref{fig:observed_and_fitted_variances}   (left) shows the {VR} values in   Equation~(\ref{eq:varratio}) for data simulated by $DW(\beta,0.7)$ and fitted by Poisson and NB distributions, respectively.  {Comparing these values of the {VR} to 1,} the plot shows cases of data overdispersed and underdispersed relative to Poisson regression. In addition, while NB regression can fit data that are overdispersed relative to Poisson regression (i.e., {VR} close to 1) well, this does not happen for underdispersed data, for which both Poisson and NB regression are inappropriate.

Figure~\ref{fig:observed_and_fitted_variances} (right) considers more closely the case of dispersion relative to Poisson regression for a range of values of $q$ and $\beta$ and shows how the DW distribution, a single distribution with as many parameters as NB regression, can capture cases of underdispersion, equidispersion and overdispersion relative to Poisson regression.

In particular, these numerical analyses have approximately shown the following:
\begin{itemize}[leftmargin=*,labelsep=5.8mm]
	\item $ 0<\beta\leq1$ is a case of overdispersion, regardless of the value of $q$.
	\item $ \beta\geq3 $ is a case of underdispersion, regardless of the value of $q$. In fact, the DW distribution approaches the Bernoulli distribution with mean $ p $ and variance $ p(1-p) $ for $ \beta \rightarrow \infty $.
	\item $1<\beta<3 $ leads to both cases of overdispersion and underdispersion depending on the value of $q$.
\end{itemize}

%%%%%%%%%%%%%%%%%%%%%%%%%%%%%%%%%%%%%%%%%%
	
\section{DW Regression Model} \label{sec:dwreg}
We now exploit this advantageous property of a DW distribution within a regression context, where the interest is to model the relationship between a count response variable and a set of covariates.

%%%%%%%%%%%%%%%%%%%%%%%%%%%%%%%%%%%%%%%%%%

\subsection{Model Formulation}
\begin{figure}[H]
	\centering
	\includegraphics[width=\linewidth]{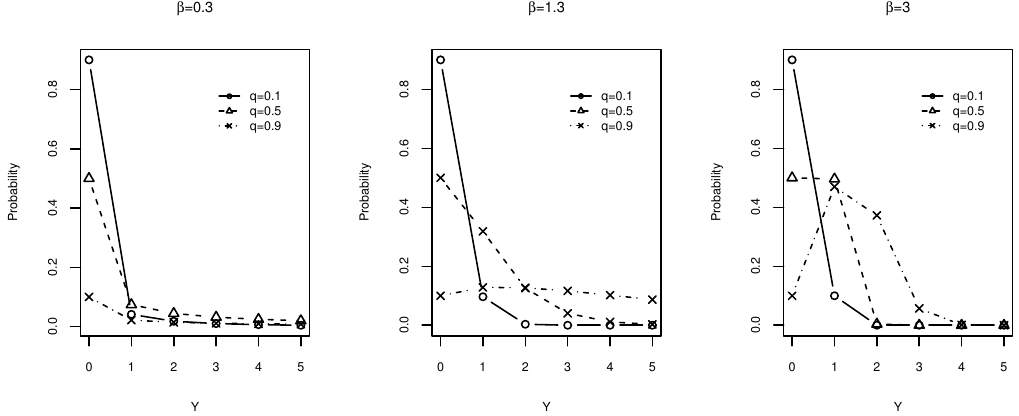}
	\caption{Effect of $ q $ on the probability mass function of the discrete Weibull (DW) distribution for different $ \beta$ values.}
	\label{fig:the effect of q on PMF}
\end{figure}

We introduce the DW regression model for count data in analogy with the continuous Weibull regression, which is well known in survival analysis and life-time modelling.  Recalling that the distribution function of a continuous Weibull distribution is given by
\[F(y;\lambda,\beta) = 1-e^{-\lambda y^{\beta}},    y\geq0,\]
with scale parameter $\lambda$, one can see that the parameter $ q $ of a DW distribution is equivalent to $ e^{-\lambda}$ in the continuous case. Because Weibull regression imposes a log link between the parameter $ \lambda $ and the predictors \citep{da2008improved,lee2003statistical}, the DW regression can be introduced via the parameter $q$.
Figure~\ref{fig:the effect of q on PMF} shows how the parameter $ q $ affects the scale and the shape of the probability mass function of the DW distribution.

From Equation~(\ref{eq:quantileDW})  {with $ \tau=\frac{1}{2} $, the median of $ Y $, denoted by $ M $, satisfies}
\begin{equation}
\label{eq:median.q}
\log\left( M+1\right) =\frac{1}{\beta} \log\big(\log(2)\big)-\frac{1}{\beta} \log\big(-\log(q)\big)
\end{equation}

Thus, in order to introduce a DW regression model, we assume that, for $ i=1, 2, \ldots, n $, the response $Y_i$ has a DW conditional distribution  $ f(y_{i},q(x_{i}),\beta|x_i) $, where $ q(x_{i}) $ is the DW parameter related to the explanatory variables $x_i$ through the link function:

\begin{equation}
\label{eq:log-log q}
\log\left( -\log(q_{i})\right)
=
\pmb{x}'_{i}\pmb{\alpha},
\qquad
\pmb{x}'_{i} \pmb{\alpha}=\alpha_{0}+ x_{i1} \alpha_{1} + \ldots +x_{iP} \alpha_{P}
\end{equation}

This transforms $ q $ from the probability scale (i.e., the interval $ [0,1] $) to the interval $ [-\infty,+\infty] $ and ensures that the parameter $ q $ remains in $ [0,1]  $.
The {$ log(-log) $} link in $q$ is also motivated by the analytical formula for the % Please check if format is ok.
quantile Equation~(\ref{eq:quantileDW}), which facilitates the interpretation of the parameters, as discussed in the next subsection. Other link functions are possible, such as the logit (analogous with geometric regression) or probit link on $q$. Moreover, the DW regression model can be introduced by relating $ \beta $ to the explanatory variable, $ f(y_{i},q,\beta(x_{i})|x_i) $, or more generally, by adding a link to both parameters, $ f(y_{i},q(x_{i}),\beta(x_{i})|x_i) $.

Then, from Equation~(\ref{eq:log-log q}), $ q_{i} $ can be expressed as
\begin{equation}	
\label{eq:q}
q_{i}=
e^{-e^{\pmb{x}'_{i} \pmb{\alpha}}}
\end{equation}
from which the conditional probability mass function of the response variable $Y_i$ given $x_i$ is as follows
\begin{equation}
\label{eq:pmf DW regression 2}
f(y_{i}|x_{i})
=
\left( e^{-e^{\pmb{x}'_{i} \pmb{\alpha}}} \right)^{y_{i}^\beta}
-
\left( e^{-e^{\pmb{x}'_{i} \pmb{\alpha}}} \right)^{(y_{i}+1)^\beta}
\end{equation}

Finally, in order to obtain the MLEs for the unknown parameters $\pmb{\alpha} $ and $ \beta $, the log-likelihood of Equation~(\ref{eq:pmf DW regression 2}) is maximised numerically using standard nonlinear optimisation tools.

%%%%%%%%%%%%%%%%%%%%%%%%%%%%%%%%%%%%%%%%%%

\subsection{Interpretation of the Regression Coefficients}
After a DW regression model has been estimated, the following can be obtained:
\begin{itemize}[leftmargin=*,labelsep=5.8mm]
	\item The fitted values for the central trend of the conditional distribution, namely, the following:
	\begin{itemize}[leftmargin=*,labelsep=5.8mm]
		\item Mean: Equation~(\ref{eq:expectedDW}), as mentioned earlier, can be calculated numerically using the approximated moments of the DW regression \citep{barbiero2013package}.
		\item Median: The quantile formula provided  in Equation~(\ref{eq:quantileDW}) can be applied. Because of the skewness, which is common for count data, the median is more appropriate than the mean. The fitted conditional median can be obtained easily from the closed-form expression of quantiles for DW regression, as
		\begin{equation}
		\label{eq:median}
		M(\pmb{x})=\Big\lceil\left(-\dfrac{\log(2)}{\log(q(\pmb{x}))}
		\right) ^{\frac{1}{\beta}}-1\Big\rceil
		\end{equation}
	\end{itemize}
	\item The conditional quantile for any $ \tau $ can be obtained from Equation~(\ref{eq:quantileDW}).
\end{itemize}

The analytical expression of the quantiles, combined with the chosen {$ log(-log) $} link function, offers a way of interpreting the parameters. Indeed, substituting Equation~(\ref{eq:q}) into Equation~(\ref{eq:median}) leads~to
\begin{equation}
\label{eq:median.intr}
\log\left( M(\pmb{x})+1\right) =\frac{1}{\beta} \log\big(\log(2)\big)-\frac{1}{\beta} \pmb{x}' \pmb{\alpha}
\end{equation}

Thus, the regression parameters $\pmb{\alpha}$ can be interpreted in relation to the log of the median. This~is in analogy with Poisson and NB models, for which the parameters are linked to the mean. In particular, $\dfrac{\log\big(\log(2) \big) - \alpha_0}{\beta} $ is related to the conditional median when all covariates are set to zero, whereas $\dfrac{-\alpha_p}{\beta} $, $ p=1, \ldots,P $, can be related to the change in the median of the response corresponding to a~one-unit change of $X_p$, keeping all other covariates constant.

\subsection{Diagnostics Checks}
After fitting a DW regression model, it is essential to consider a diagnostics analysis to investigate the appropriateness of the model. Given that the response is discrete, we advise performing a residual analysis on the basis of the randomised quantile residuals, as developed by \citep{dunn1996randomized} and used in many other studies (e.g., \citep{ospina2012general,vanegas2013diagnostic}). In particular, we let
\begin{equation}
\label{eq:randomizesquantileresiduals}
r_i=\Phi^{-1}(u_i) \qquad \qquad i=1, \ldots, n
\end{equation}
where $ \Phi(.) $ is the standard normal distribution function and $ u_i $ is a uniform random variable on the~interval
\begin{equation*}
\begin{split}
\left( a_i, b_i \right] & = \bigg( \underset{y \uparrow y_i}{\mathrm{lim}}
F(y;\hat{q_i},\hat{\beta}), F(y_i;\hat{q_i},\hat{\beta}) \bigg] \\
& \approx \bigg[ F(y_i -1;\hat{q_i},\hat{\beta}),  F(y_i;\hat{q_i},\hat{\beta})\bigg]
\end{split}
\end{equation*}

These residuals  follow the standard normal distribution, apart from the sampling variability in $\hat{q_i}$ and $\hat{\beta}$. Hence, the validity of a DW model can be assessed using goodness-of-fit investigations of the normality of the residuals, such as {Q-Q} plots and normality tests. Simulated envelopes can be added to the {Q-Q} plots, as in \citep{ferrari2004beta,garay2011estimation,saez2013hyper,atkinson1985plots}. 	 % Please confirm

In addition to the residual analysis, it is informative also to check whether the data shows any underdispersion or overdispersion relative to the specified DW conditional distribution. In the case of good fitting, we would expect the ratio of observed and theoretical variance in Equation~(\ref{eq:varratio}) to be close to 1 for each $x$. We expand more on this point in the next section.

%%%%%%%%%%%%%%%%%%%%%%%%%%%%%%%%%%%%%%%%%%

\section{DW Regression Naturally Handles Covariate-Specific Dispersion} \label{sec:natuhan}
We have already shown in Section \ref{sec:disp} how DW regression can model both data that are overdispersion and underdispersed relative to Poisson regression. In this section, we investigate this further within a regression context. Here, it is also possible that the conditional variance is larger than the conditional mean for a specific covariate pattern (overdispersion), but the conditional variance is smaller than the conditional mean for another covariate pattern (underdispersion).

In the literature, regression models for count data that can capture underdispersion or both types of overdispersion and underdispersion simultaneously are in the form of extended versions of Poisson regression, such as quasi-Poisson, COM--Poisson or hyper-poisson regression \citep{saez2013hyper}. In the case of mixed types of dispersion, the dispersion parameter can be assumed to be linked to the covariates. However, a covariate-dependent dispersion increases the complexity of the model significantly and reduces its interpretability. Thus, in practice, most implementations fix the dispersion parameter and assume that only the mean is linked to the covariates. As the DW distribution naturally accounts for overdispersion and underdispersion, a DW regression model becomes a simple and attractive alternative to existing regression models for count data.

 {We emphasize this point by a simple simulation study. We have considered two cases, a small sample size ($n=25$) and a large sample size ($ n=600 $),} with two covariates, $X_1 \sim N(0,1)$ and $X_2 \sim \rm{Uniform}(0,10)$. We assumed the regression parameters to take values $ \pmb{\alpha}=(\alpha_{0},\alpha_{1},\alpha_2)=(0.5,0.4,-0.3)$. In addition, the parameter $ \beta $ of the DW regression was assumed to be $ \beta=2.1 $. Then, for each case, we respectively sampled 25 and 600 values of the covariates and the corresponding response from $ DW(q_i,\beta) $, where $ q_i $ is calculated as in Equation~(\ref{eq:q}), for $ i=1, \ldots, n $. Table~\ref{tab:Simulation2results} reports the estimates of the parameters, together with the average bias and the mean-squared error (MSE) over $1000$ iterations.
\begin{table}[H]
	\caption{Simulation study: discrete Weibull (DW) parameter estimates by maximum likelihood estimators (MLEs), together with bias, mean-squared error (MSE) and length of the 95\% confidence interva (Cl), averaged over 1000 iterations.}
	\label{tab:Simulation2results}
%	\begin{center}
\centering
		\begin{tabular}{cccccc}
			\toprule
			 &  & \bf{MLE} & \bf Bias & \bf MSE & \bf 95\% \bf CI \bf Length \\
			\midrule
			\multirow{4}{*}{$n=25$}
			& $ \alpha_{0} $ & 0.6467 & 0.1467 & 0.2932  & 1.7324  \\
			& $ \alpha_{1} $ & 0.4908 & 0.0908 & 0.0763  & 0.8821  \\
			& $ \alpha_{2} $ & $-$0.3651 & $-$0.0651 & 0.0241  & 0.4694  \\
			& $ \beta $ & 2.5924  & 0.4924 & 0.8455  & 2.4718  \\
			\midrule
			\multirow{4}{*}{$n=600$}
			& $ \alpha_{0} $ & 0.5074 & 0.0074 & 0.009  & 0.3611  \\
			& $ \alpha_{1} $ & 0.402 & 0.002 & 0.0021  & 0.1782  \\
			& $ \alpha_{2} $ & $-$0.3033 & $-$0.0033 & 0.0004  & 0.0789  \\
			& $ \beta $ & 2.1196  & 0.0196 & 0.0086  & 0.3549  \\
			\bottomrule
		\end{tabular}
%	\end{center}
\end{table}

\begin{figure}[H]
	\centering
	\includegraphics[scale=0.729]{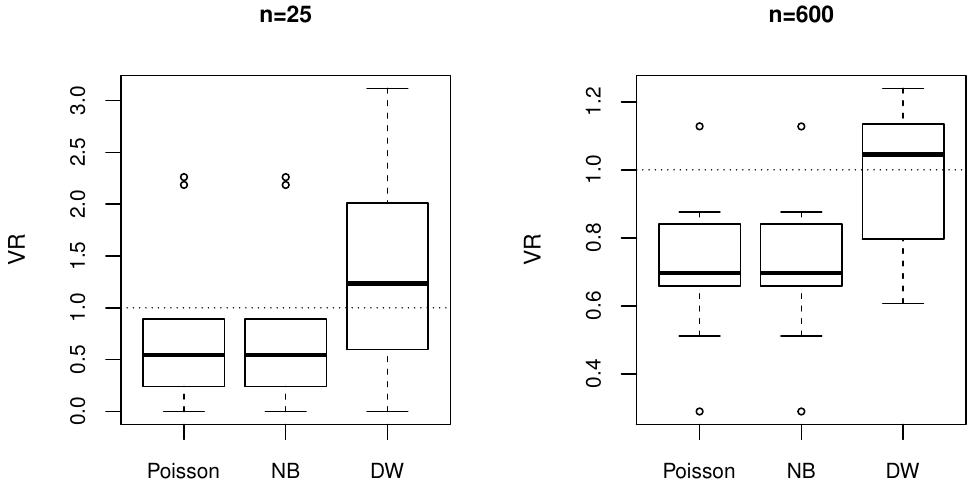}
	\caption{Distribution of ratios of observed and theoretical conditional variance on simulated data from the discrete Weibull (DW) regression model, with the theoretical variance fitted by Poisson, negative binomial (NB) and DW models.}
	\label{fig:VR_sim}
\end{figure}

Figure~\ref{fig:VR_sim} shows a  {boxplot} of the dispersions in Equation~(\ref{eq:varratio}) in the case of Poisson, NB and DW fitting. A note is required on the calculation of these ratios, as the observed variance could not be computed for each individual covariate vector $x$. For the calculation, we split the response values into 10 groups of similar size, on the basis of the percentiles of the linear predictors $ \pmb{x}'\pmb{\alpha}$. Then the observed variance was computed within each group, while the theoretical variance was averaged within each group. If the model was well specified, we would expect these values to have been close to 1. This is shown in Figure~\ref{fig:VR_sim} for DW regression, which was the model used in the simulation. Poisson and NB regression showed underdispersion in most cases and overdispersion in two cases. Thus, this simulation shows a simple scenario of a mixed level of dispersion, which could not be captured well by standard Poisson and NB models.

%%%%%%%%%%%%%%%%%%%%%%%%%%%%%%%%%%%%%%%%%%

\section{Application to Real Datasets} \label{sec:realdata}
To demonstrate the ability of the DW regression model to handle overdispersion and underdispersion automatically, in this section, DW regression has been applied to different datasets that show various types of dispersions relative to Poisson regression.
The first subsection uses an underdispersed dataset, while the second uses an overdispersed case. The third subsection focuses on  a zero-inflated dataset. Finally,  an illustrative example for the mixed level of dispersion is provided.  Various popular count data regression models, namely, Poisson regression (R function {\em glm}), NB regression (R function {\em glm.nb}), COM--Poisson regression (R package {\em COMPoissonReg} \citep{COMPoissonReg}), GP regression (R package {\em VGAM} \citep{yee2010vgam}), and zero-inflated and hurdle models ({\em pscl} R package \citep{zeileis08}), have been applied and compared with DW regression by means of classical {AIC and BIC} criteria \citep{dayton2003model}. % Please define.

%%%%%%%%%%%%%%%%%%%%%%%%%%%%%%%%%%%%%%%%%%

\subsection{The Case of Underdispersion: Inhaler Usage Data}

For this example, we used data from  \cite{grunwald2011statistical}. These consisted of 5209  observations  and reported the daily count of using (albuterol) asthma inhalers for 48 children, aged between $6$ and $13$, suffering from asthma during the school day, for a period of time at the Kunsberg School of National Jewish Health in Denver, Colorado. The main objective of this analysis was to investigate the relationship between the inhaler use (representing the asthma severity) and air pollution, which was recorded by four covariates: the percentage of humidity, the barometric pressure (in mmHG/1000), the average daily temperature (in $^\circ$F/100), and the morning levels of $PM_{25}$, which are small air particles less than 25 mm in diameter. The response variable, which was the inhaler use count, had a sample mean of $ 1.2705 $ and variance of $ 0.8433 $, thus pointing to a case of underdispersion relative to Poisson regression.

The results in Table~\ref{tab:MLEs_example2} suggest that DW and COM--Poisson regression provided better fitting than both Poisson and NB models, according to both AIC and BIC. 
 {The COM--Poisson regression considered here was based on} \citep{sellers2008flexible} {with the following probability mass function}:
\begin{equation*}
	P(Y_i=y_i)=\frac{{\lambda_i}^{y_i}}{(y_i!)Z(\lambda_i,\nu)}  \qquad \qquad Z(\lambda_i,\nu)=\sum_{s=0}^{\infty} \frac{{\lambda_i}^{s}}{(s!)^{\nu}}
\end{equation*}

\begin{figure}[H]
	\centering
	\includegraphics[width=0.54\linewidth]{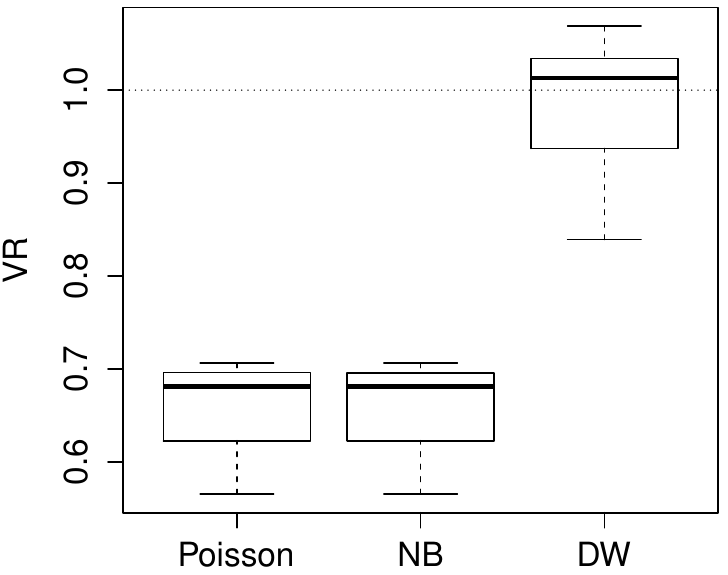}
	\includegraphics[width=0.42\linewidth]{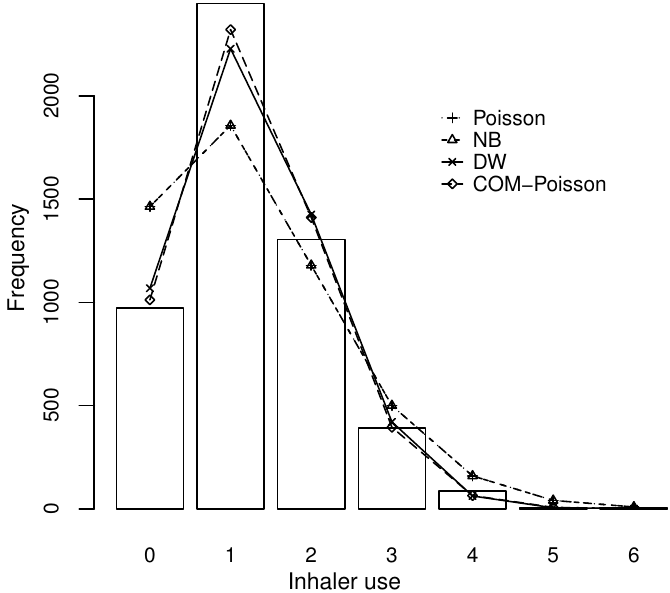}
	\caption{Comparison of discrete Weibull (DW) regression with  {the other regression models} on inhaler use data. {\bf Left}: distribution of ratios of observed and theoretical conditional variance on the  data fitted by Poisson, negative binomial (NB) and DW regression, respectively. {\bf Right}: observed and expected frequencies for each model.}
	\label{fig:VR_2}
\end{figure}

\begin{figure}[H]
	\centering
	\includegraphics[scale=0.784]{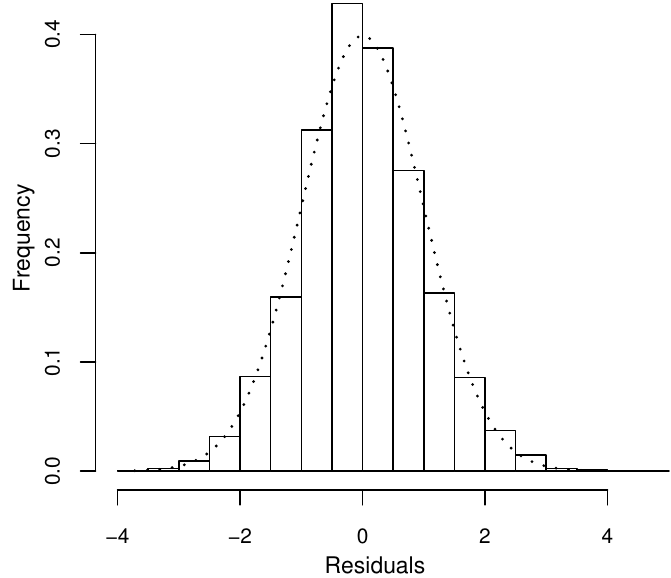}
	\caption{Residuals analysis for inhaler use data for the discrete Weibull (DW) regression model: histogram of randomised quantile residuals with superimposed {$N(0,1)$} density (dotted line).}
	\label{fig:residuals_hist_ex2} % Italic, please confirm.
\end{figure}

 GP regression was also attempted on this dataset, but it did not improve on COM--Poisson regression (AIC = 13550.17) and is thus not reported in the table. For DW regression, the parameters are reported with the parameterisation linked to the median, as previously described. The left panel of Figure~\ref{fig:VR_2} indicates underdispersion relative to Poisson and NB regression across the full range of the covariates and a good fit of DW regression compared to the other models ({VR} values close to 1). COM--Poisson regression could not be added to this plot because of the complexity of calculating the theoretical variances in this case. The right panel compares the observed and expected frequencies for the  {four} models and shows again a good fit for DW regression. Finally, Figure~\ref{fig:residuals_hist_ex2} plots the randomised quantile residuals from the DW regression model, which only moderately depart from normality ($p$-value of Kolmogorov--Smirnov (KS) test: 0.025).

\begin{table}[H]
	\centering
	\captionof{table}{Maximum likelihood estimates, {AIC and BIC} from different regression models fitted to  the inhaler use data.}
	\label{tab:MLEs_example2}
	\resizebox{\linewidth}{!}{
		\begin{tabular}{ccccclccc}
			\toprule
			{} & \bf Humidity & \bf Pressure & \bf Temperature & \bf Particles & \bf Other & \bf  AIC & \bf BIC
			\\
			\midrule
			Poisson &  $-$0.1125  & 4.0950 & $-$0.2035 & 0.0225 &  --- & 13915.47 & 13948.26  \\
			{NB} & $-$0.1125 &4.0950& $-$0.2035 & 0.0225 &  $ \hat{k} =31905.28$  & 13917.54 & 13956.89 \\
			{COM}--Poisson & $-$0.1724 & 6.2864 & $-$0.3128 & 0.0348 & $  \hat{\nu} =1.9203$& 13450.77 & 13490.12\\				
			{DW} &$-$0.1050 & 2.6376 & $-$0.1735  & 0.0136   & $  \hat{\beta} =2.1277$ & 13484.36 & 13523.71 \\
			\bottomrule
	\end{tabular}}
\end{table} % Please define all abbreviations at first use in table.

\vspace{-12pt}

%%%%%%%%%%%%%%%%%%%%%%%%%%%%%%%%%%%%%%%%%%

\subsection{The Case of Overdispersion: Doctor Visits from German Health Survey Data}
 {This dataset comes from the German Health Survey} and is available in the {\em COUNT} R package~\citep{Hilbe2014count}, under the name of {\em badhealth}. The response variable is the number of visits to certain doctors during 1998. Two predictors are considered: an indicator variable representing patients claiming to be in bad health  (1) or not (0), and the age of the patient. The response variable  ranges from $0$ to $40$ visits and has a sample mean of $ 2.3532 $ and variance of $ 11.9818 $, suggesting overdispersion relative to Poisson regression.
Indeed, a comparison of Poisson and NB distributions solely on the response variable using a likelihood ratio test ({\em lmtest} R package \citep{lmtest}) shows evidence of overdispersion with a chi-square test statistic of $ 1165.267 $ and $p$-value $ <0.001$.

\begin{figure}[H]
	\centering
	\includegraphics[width=0.41\linewidth]{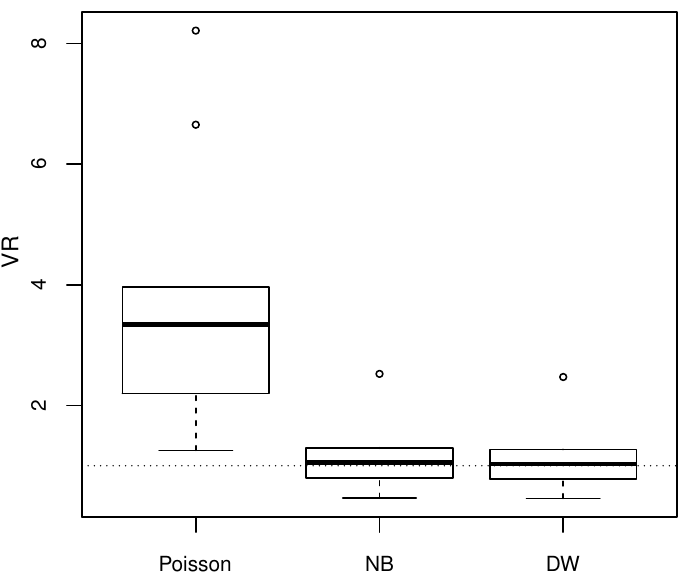}
	\includegraphics[width=0.57\linewidth]{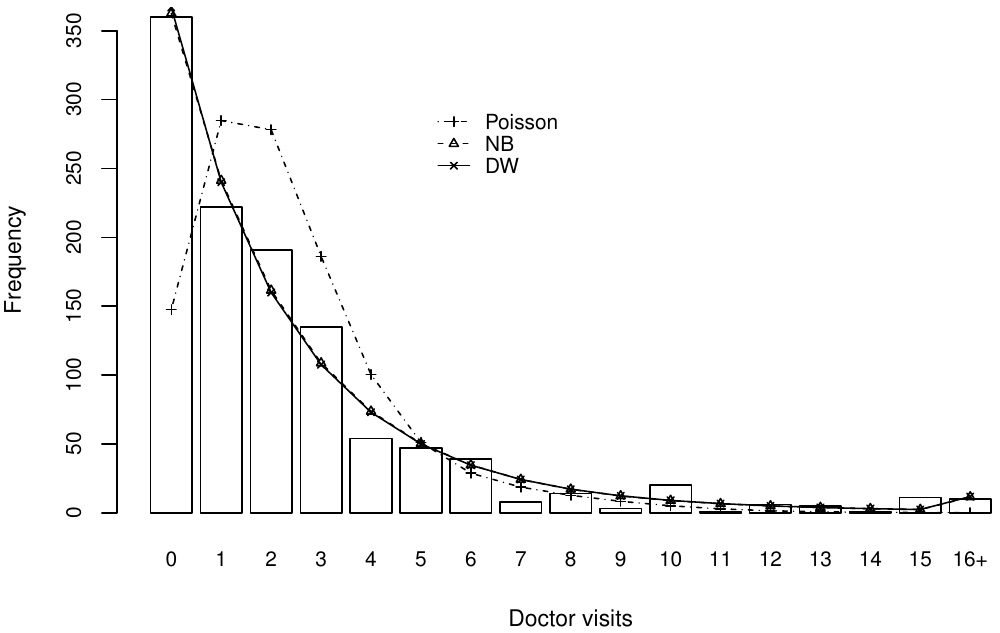}
	\caption{Comparison of discrete Weibull (DW) with negative binomial (NB) and Poisson regression on doctor visits from German Health Survey data. {\bf Left}: distribution of ratios of observed and theoretical conditional variance on the data fitted by Poisson, NB and DW regression, respectively. {\bf Right}: observed and expected frequencies for each model.}
	\label{fig:VR_3}
\end{figure}

\begin{figure}[H]
	\centering
	\includegraphics[scale=0.67]{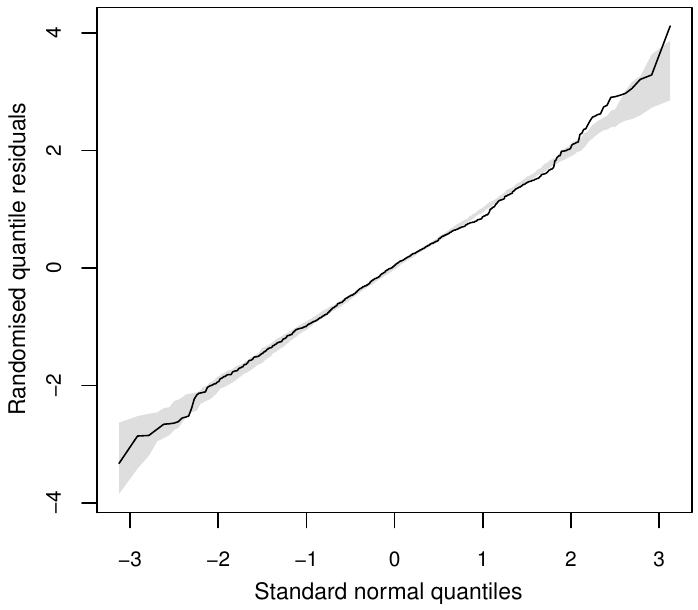}
	\caption{Residuals analysis for doctor visits from German Health Survey data: {Q-Q} %Please confirm.
	 plot of randomised quantile residuals of the discrete Weibull (DW) regression model. }
	\label{fig:residuals_simenvelope_ex3}
\end{figure}

After fitting three regression models and comparing them via AIC and BIC, Table~\ref{tab:MLEs_example3} shows that the DW model was only marginally superior to the NB, but both DW and NB regression models gave much  better fits to the data than the Poisson regression model. The left panel of Figure~\ref{fig:VR_3} indicates a case of overdispersion relative to Poisson regression across the whole range of the covariates. Additionally, it~indicates the better fit of the NB and DW models than Poisson regression with {VR} values closer to $ 1 $. The right panel confirms the good fit of NB and DW regression. For visualisation purposes, the small number of observations larger than 16 are grouped together in this plot. Finally, Figure~\ref{fig:residuals_simenvelope_ex3} shows that the residuals closely followed a normal distribution (KS $p$-value: 0.06), with not many points falling outside the simulated 95\% envelope's bounds.
\begin{table}[H]
	\centering
	\captionof{table}{Maximum likelihood estimates, {AIC and BIC} from different regression models fitted to the doctor visits from German Health Survey data.}
	\label{tab:MLEs_example3}
	\begin{tabular}{cccccc}
		\toprule
		{} & \bf Bad Health &\bf  Age & \bf Other & \bf AIC & \bf BIC
		\\
		\midrule	
		Poisson & 1.1083  &  0.0058 & ---  & 5638.552 & 5653.634 \\
		{NB} & 1.1073 & 0.0070 &  $ \hat{k} =0.9975$ & 4475.285  &  4495.394 \\
		{DW} & 1.0068 & 0.0120 & $  \hat{\beta} =0.9887$  & 4474.973 & 4495.083  \\
		\bottomrule
	\end{tabular} 
\end{table}

\vspace{-12pt}

\vspace{-12pt}

%%%%%%%%%%%%%%%%%%%%%%%%%%%%%%%%%%%%%%%%%%

\subsection{The Case of Excessive Zeros: Doctor Visits from the United States Data}
The following dataset illustrates the case of excessive zero counts. Thus, besides the Poisson, NB, and DW regression, we include also zero-inflated and hurdle models in the comparison.  {Particularly, zero-inflated Poisson (ZIP), zero-inflated negative binomial (ZINB), hurdle Poisson (HP) and hurdle negative binomial (HNB) are considered} with the logit link function for the binomial distribution representing the probability of the extra zeros (R package {\em pscl} \citep{zeileis08}). The data are available from the {\em Ecdat} R package, under the name {\em Doctor}. The data consist of 485 observations from the United States in the year 1986 and contain four variables for each patient: the number of doctor visits, which is taken as the response, and the number of children in the household,  a measure of access to health care and a measure of health status (larger positive numbers are associated with poorer health). The response variable in this study, the number of doctor visits, has approximately $50\%$ of zeros, and thus it could be considered as a case of excessive zeros. Indeed, the response variable has a mean of $  1.6103 $ and variance of $ 11.2011$, and the likelihood ratio test between NB and Poisson regression returned a test statistic of $ 599.61 $ and a $p$-value of $ <0.001 $.

\begin{figure}[H]
	\centering
	\includegraphics[width=0.41\linewidth]{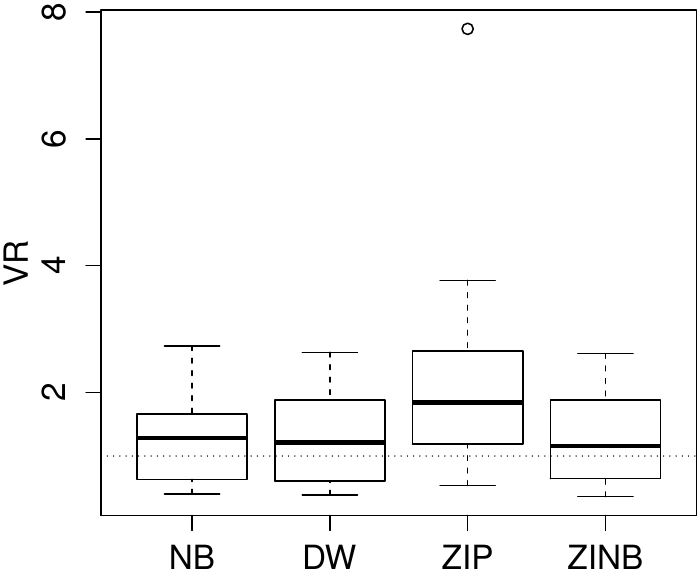}
	\includegraphics[width=0.56\linewidth]{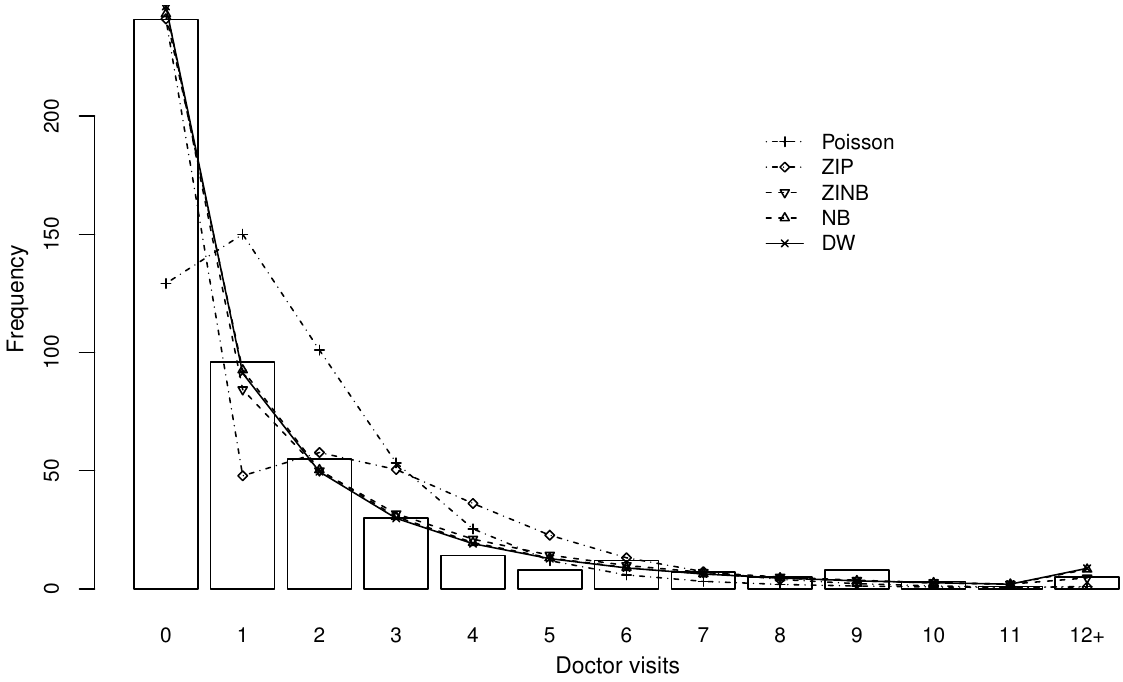}
	\caption{Comparison of discrete Weibull (DW) regression with negative binomial (NB), zero-inflated Poisson (ZIP) and zero-inflated negative binomial (ZINB) regression on doctor visits from the United States data. Left: distribution of ratios of observed and theoretical conditional variance on the data. Right: observed and expected frequencies for each model.}
	\label{fig:VR_5}
\end{figure}

Table~\ref{tab:MLEs_example5} shows the best fit for the DW regression model in terms of AIC and BIC. The left panel of  Figure~\ref{fig:VR_5}  shows a case of overdispersion relative to Poisson regression across the full range of covariates and a good fit for DW and ZINB regression. We excluded Poisson regression from this plot, as the {VR} values were very large in this case,  {as well as the hurdle models, as they provided almost identical results to the corresponding zero-inflated models}.
The right panel confirms the good fit of ZINB and DW regression. For visualisation purposes, the small number of observations larger than 12 are grouped together on this plot.
As in the previous example, the residuals of the DW model were well approximated by a normal distribution (KS $p$-value: 0.856). This example shows how DW regression in its simplest form can also model cases of excessive zeros, although additional zero-inflated components could also be added to a DW model if necessary and will be explored in future work.
\begin{table}[H]
	\centering
	\captionof{table}{Maximum likelihood estimates, {AIC and BIC} from different regression models fitted to doctor visits from the United States data.} \label{tab:MLEs_example5}
	\scalebox{0.859}[0.859]{
	\resizebox{\linewidth}{!}{
		\begin{tabular}{ccccccccc}
			\toprule
			& \bf Children & \bf  Access & \bf Health  & \bf Other & \bf AIC & \bf BIC \\
			\midrule
			Poisson & $-$0.1759  & 0.9369 & 0.2898  & --- & 2179.487 &  2196.223 \\	
			%%%%					
			{NB} & $-$0.1706  & 0.4197  & 0.3154  & $ \hat{k} =0.5525$ & 1581.88 & 1602.801 \\
			\midrule
		{Zero-inflated models} 		&&&&&&&&\\\midrule
			Poisson 		&&&&&&&&\\				
			%%%%
			Count model  &  $-$0.1498 & 0.8053 & 0.1736  & --- & \multirow{2}{*}{1885.813} &\multirow{2}{*}{1919.287}   \\
			Logit model &  0.0843 & $-$0.1048 & $-$0.4147  & --- & &\\
			NB 		&&&&&&&&\\									%%%%
			Count model  &$-$0.1414  & 0.6491 & 0.2239 &  $ \hat{k}=0.6869$   & \multirow{2}{*}{1578.5} &  \multirow{2}{*}{1616.158}   \\
			Logit model & 0.2465 & 1.2085  & $-$2.0676  & --- & &\\
			\midrule
			{Hurdle models} 		&&&&&&&&\\\midrule				
			%%%%
			Logit model &$-$0.1462  & 0.4252  &0.4524 & --- & --- & ---\\
			Poisson count model & $-$0.1506  & 0.8143 & 0.1733  & --- & 1885.808 & 1919.281\\
			%%%%
			NB count model & $-$0.1664  & 0.5404& 0.2157  & $ \hat{k} =0.2596$ & 1576.302 &  1613.959  \\
			\midrule
			%%%%
			{DW} & $-$0.1309 & 0.3403 &0.2758  & $  \hat{\beta} =0.7823$& 1575.796 & 1596.717 \\
			\bottomrule
	\end{tabular}}}
\end{table}

%%%%%%%%%%%%%%%%%%%%%%%%%%%%%%%%%%%%%%%%%%

\subsection{The Case of a Mixed Level of Dispersion: Bids Data}
In this section, we report the analysis of a dataset for which a mixed level of dispersion was observed; that is, the conditional distribution is overdispersed relative to Poisson regression for some covariate pattern but is underdispersed for another covariate pattern. The data are taken from \cite{cameronaa1997count} and are available in the  {\em Ecdat} R package under the name of {\em Bids}.  The data record the number of bids received by $ 126 $ U.S. firms that were targets of tender offers during a certain period of time. The dependent variable here is the number of bids, with a mean of $ 1.7381 $ and a variance of $ 2.0509 $. The objective of the study was to investigate the effect of some variables on the number of bids. For this analysis, we considered the following covariates: bid price, taken as the price at a particular week divided by the price 14 working days before the bid; the size, that ism the total book value of assets measured in billions dollars; and a regulator, a dummy variable, which was 1 if there was an intervention by federal regulators and 0 otherwise.

\begin{figure}[H]
	\centering
	\includegraphics[scale=0.56]{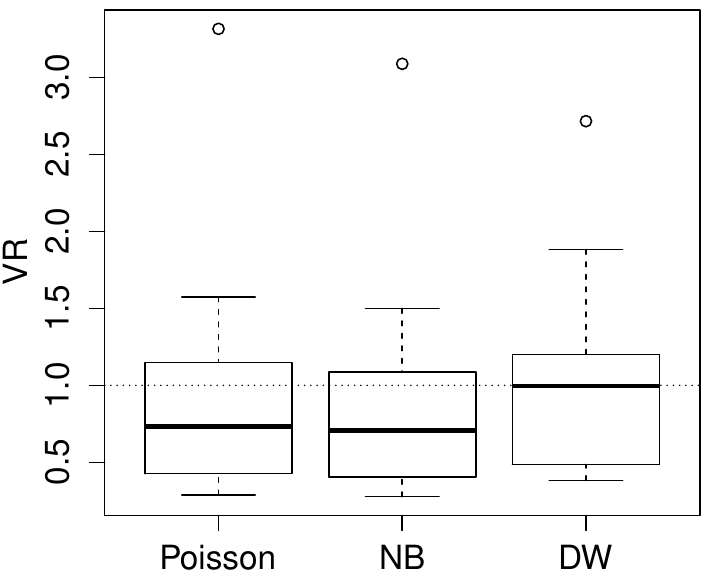}
	\caption{Distribution of ratios of observed and theoretical conditional variance on the Bids data fitted by Poisson, negative binomial (NB) and discrete Weibull (DW) regression.}
	\label{fig:VR_7}
\end{figure}

Figure~\ref{fig:VR_7} and Table~\ref{tab:MLEs_example7} show once again a very good fit of the DW regression model to these data, compared to Poisson and NB regression. Figure~\ref{fig:VR_7} shows a mixed level of dispersion relative to Poisson and NB regression, with most covariate patterns leading to underdispersion, but with a small number of overdispersed cases. The DW model has a clearer distribution of {VR} values around 1, and it also fits the data well, with a KS $p$-value of 0.11 for the randomised quantile residuals.
\begin{table}[H]
	\centering
	\captionof{table}{Maximum likelihood estimates, AIC and BIC from different regression models fitted to Bids~data.}
	\label{tab:MLEs_example7}
	\begin{tabular}{cccccccc}
		\toprule
		{} & \bf Price &\bf  Size & \bf Regulator  &\bf  Other & \bf AIC & \bf BIC \\
		\midrule
		Poisson &  $-$0.7849 & 0.0362  &0.0547   & ---  & 402.2602  & 413.6054  \\
		{NB} &  $-$0.7824  & 0.0369  &  0.0544  & $ \hat{k} =33.3289$  & 403.9481 & 418.1295 \\
		{DW} &  $-$0.6761  & 0.0552  & 0.0293 & $  \hat{\beta} =1.9403$   & 395.1214 & 409.3028 \\
		\bottomrule
	\end{tabular}
\end{table}
\vspace{-12pt}

%%%%%%%%%%%%%%%%%%%%%%%%%%%%%%%%%%%%%%%%%%
	
\section{Conclusions}

In this paper, we introduce a regression model based on a DW distribution and show how this model can be seen as a simple and unified model to capture different levels of dispersion in the data, namely, underdispersion and overdispersion relative to Poisson regression. This is an attractive feature of DW regression, similar to the flexibility of the continuous Weibull distribution in adapting to a variety of hazard rates. In addition, the proposed DW regression model, unlike generalised linear models in which the conditional mean is central to the interpretation, has the advantage that the conditional quantiles can be easily extracted from the fitted model and the regression coefficients can be easily interpreted in terms of changes in the conditional median. This is particularly useful, as most count data have a highly skewed distribution.

A popular model for underdispersion is the COM--Poisson regression model. However, the probability mass function of COM--Poisson regression is not in
a closed form and it contains an infinite sum, which requires an approximate computation. In fact, the COM--Poisson implementation that was used for the examples in this paper requires more computational time than the DW regression model, which uses a straightforward MLE procedure. This is particularly beneficial in the case of large sample sizes.  While NB regression is the most widely applied model for overdispersion, the DW regression model is shown  to be an attractive alternative to the NB regression model for overdispersion.

The DW regression model described in this paper is implemented in the R package {\em DWreg}, freely available in CRAN \cite{DWreg2016Response}.

%%%%%%%%%%%%%%%%%%%%%%%%%%%%%%%%%%%%%%%%%%
\vspace{6pt}
%%%%%%%%%%%%%%%%%%%%%%%%%%%%%%%%%%%%%%%%%%
\acknowledgments{This work was supported by the Major Program of Hubei Provincial Department of Education for Philosophy and Social Science Research  (Grant No. 17ZD018 ) and the National Institute for Health Research Method Grant (NIHR-RMOFS-2013-03-
09) .} %Please add

%Hadeel S. Klakattawi, Veronica Vinciotti  and Keming Yu

\authorcontributions{{Under the idea and detailed guidance of the corresponding author, Hadeel S. Klakattawi together with Veronica Vinciotti explored details of inference. Hadeel S. Klakattaw carried out all
numerical calculations and a first draft, and Veronica Vinciotti and the Keming Yu polished the writing. All authors have read and approved the final manuscript.}} %Please add

\conflictsofinterest{{The authors declare no conflict of interest.}} %Please  confirm.

\abbreviations{The following abbreviations are used in this manuscript:\\

	\noindent
	\begin{tabular}{@{}ll}
		DW & Discrete Weibull\\
		NB & Negative binomial\\
		GP & Generalised Poisson \\
		ZIP & Zero-inflated Poisson\\
		ZINB & Zero-inflated negative binomial\\
		HP & Hurdle Poisson\\
		ZINB & Hurdle negative binomial\\
		MSE & Mean-squared error\\
		MLE & Maximum likelihood estimator\\
		VR & Variance ratio\\
		KS & Kolmogorov--Smirnov
\end{tabular}}
\vspace{-6pt}

\reftitle{References}

%%%%%%%%%%%%%%%%%%%%%%%%%%%%%%%%%%%%%%%%%%
\end{document}